\documentclass[doublecol]{epl2} 
\title{Enhanced conduction band density of states in intermetallic EuTSi$_3$ (T=Rh, Ir)}
\shorttitle{Enhanced conduction band density of states in EuTSi$_3$} 

\author{A. Maurya\inst{1} \and P.Bonville\inst{2} \and A. Thamizhavel\inst{1} \and S. K. Dhar\inst{1}}
\shortauthor{A. Maurya \etal}

\institute{                    
\inst{1} Department of Condensed Matter Physics and Materials
Science, Tata Institute of Fundamental Research, Homi Bhabha Road,
Colaba, Mumbai 400 005, India\\
\inst{2} CEA, Centre d'Etudes de Saclay, DSM/IRAMIS/Service de Physique de l'Etat Condens\'e and CNRS UMR 3680, 91191 Gif-sur-Yvette, France
}

\newcommand{\eis}{EuIrSi$_3$}
\newcommand{\ers}{EuRhSi$_3$}
\newcommand{\mub}{$\mu_{\rm B}$}

\abstract{
We report on the physical properties of single crystalline \ers\ and polycrystalline \eis, inferred from magnetisation, electrical transport, heat capacity and $^{151}$Eu M\"ossbauer spectroscopy. These previously known compounds crystallise in the tetragonal BaNiSn$_3$-type structure. The single crystal magnetisation in \ers\ has a strongly anisotropic behaviour at 2\,K with a spin-flop field of 13\,T, and we present a model of these magnetic properties which allows the exchange constants to be determined. In both compounds, specific heat shows the presence of a cascade of two close transitions near 50\,K, and the $^{151}$Eu M\"ossbauer spectra demonstrate that the intermediate phase has an incommensurate amplitude modulated structure. We find anomalously large values, with respect to other members of the series, for the RKKY N\'eel temperature, for the spin-flop field (13\,T), for the spin-wave gap ($\simeq$ 20-25\,K) inferred from both resistivity and specific heat data, for the spin-disorder resistivity in \ers\ ($\simeq 35$\,$\mu$Ohm.cm) and for the saturated hyperfine field (52\,T). We show that all these quantities depend on the electronic density of states at the Fermi level, implying that the latter must be strongly enhanced in these two materials. EuIrSi$_3$ exhibits a giant magnetoresistance ratio, with values exceeding 600\% at 2\,K in a field of 14\,T.}

\begin{document}

\maketitle

\maketitle

\section{Introduction}

Divalent-Eu intermetallic compounds order magnetically due to the indirect RKKY exchange interaction \cite{freeman} between the Eu 4f-spins. Several Eu-based compounds with composition EuTX$_3$, where T is a $d$-transition element and X~=~Si or Ge, crystallizing in the non-centrosymmetric BaNiSn$_3$-type structure are known. Of these, the magnetic properties of EuPtSi$_3$ \cite{Neeraj2010EuPtSi3}, EuPtGe$_3$ \cite{Neeraj2012EuPtGe3}, EuPdGe$_3$ \cite{EuPdGe3} and EuNiGe$_3$ \cite{Johnston_EuNiGe3,Arvind_EuNiGe3} have recently been reported in the literature in single crystal samples. In these materials, an anisotropic behaviour of the 2\,K magnetisation seems to be the prerequisite for the existence of a cascade of close transitions, around 15\,K: a transition from the paramagnetic to an incommensurate, moment modulated antiferromagnetic (AF) state occurs first, followed by another one to a single moment regular AF state, a few K below. This is the case for EuPtSi$_3$ and EuNiGe$_3$, whereas EuPtGe$_3$ shows a unique transition and an isotropic behaviour of the magnetisation.

Here, we report on the magnetic properties of iso-structural EuTSi$_3$ (T=Rh and Ir) compounds, with a single crystal sample for \ers\ only. An early M\"ossbauer spectroscopy study of these two compounds was performed in Ref.\cite{Chevalier}. We show that these two materials belong to the ``transition cascade'' type, with an anisotropic behaviour of the magnetisation documented for \ers. We present in addition specific heat, transport and $^{151}$Eu M\"ossbauer spectroscopy data. We find that the magnetic and transport properties in these two materials are notably enhanced with respect to those in other members of the series, and we show that this enhancement can be attributed to an unusually large density of conduction electron states at the Fermi level $n(E_F)$. As a remarkable result, the values of the spin-wave gap derived from such different techniques as resistivity, specific heat and single crystal magnetisation measurements are in good agreement.

\section{Experimental}
Polycrystalline samples of \eis\ and \ers\ were prepared by melting Eu (99.9\%  purity), Ir/Rh (99.99\%) and Si (99.999\%) in an arc furnace under an inert argon atmosphere.  
Single crystal growth of the two Eu compounds was tried using Sn and In as flux and following the same protocol as reported in Refs.\cite{Neeraj2010EuPtSi3,Neeraj2012EuPtGe3}. Powder-diffraction spectra were recorded on a Phillips Pan-analytical set up using Cu-$K_\alpha$ radiation. The magnetisation as a function of field (up to 16\,T) and temperature (1.8 to 300\,K) was measured using Quantum Design MPMS and VSM magnetometers. The electrical resistivity between 1.8 and 300\,K in zero and applied fields, and the heat capacity were measured in a Quantum Design PPMS set-up. $^{151}$Eu M\"ossbauer spectra were recorded at various temperatures using a commercial $^{151}$SmF$_3$ source mounted on a constant acceleration spectrometer. 
\section{Results and Discussion}

\subsection{Structure}
Our attempt to grow single crystals succeeded only for \ers, with In as flux. Powder diffraction spectra of \eis\ and \ers\ could be indexed on the basis of the BaNiSn$_3$ type structure (space group $I4mm$). The lattice parameters obtained by the Rietveld analysis of the powder diffraction spectra are in good agreement with the previously reported values \cite{Chevalier}. 

\subsection{Susceptibility and isothermal magnetisation}
%
\begin{figure}[h]
\includegraphics[width=0.40\textwidth]{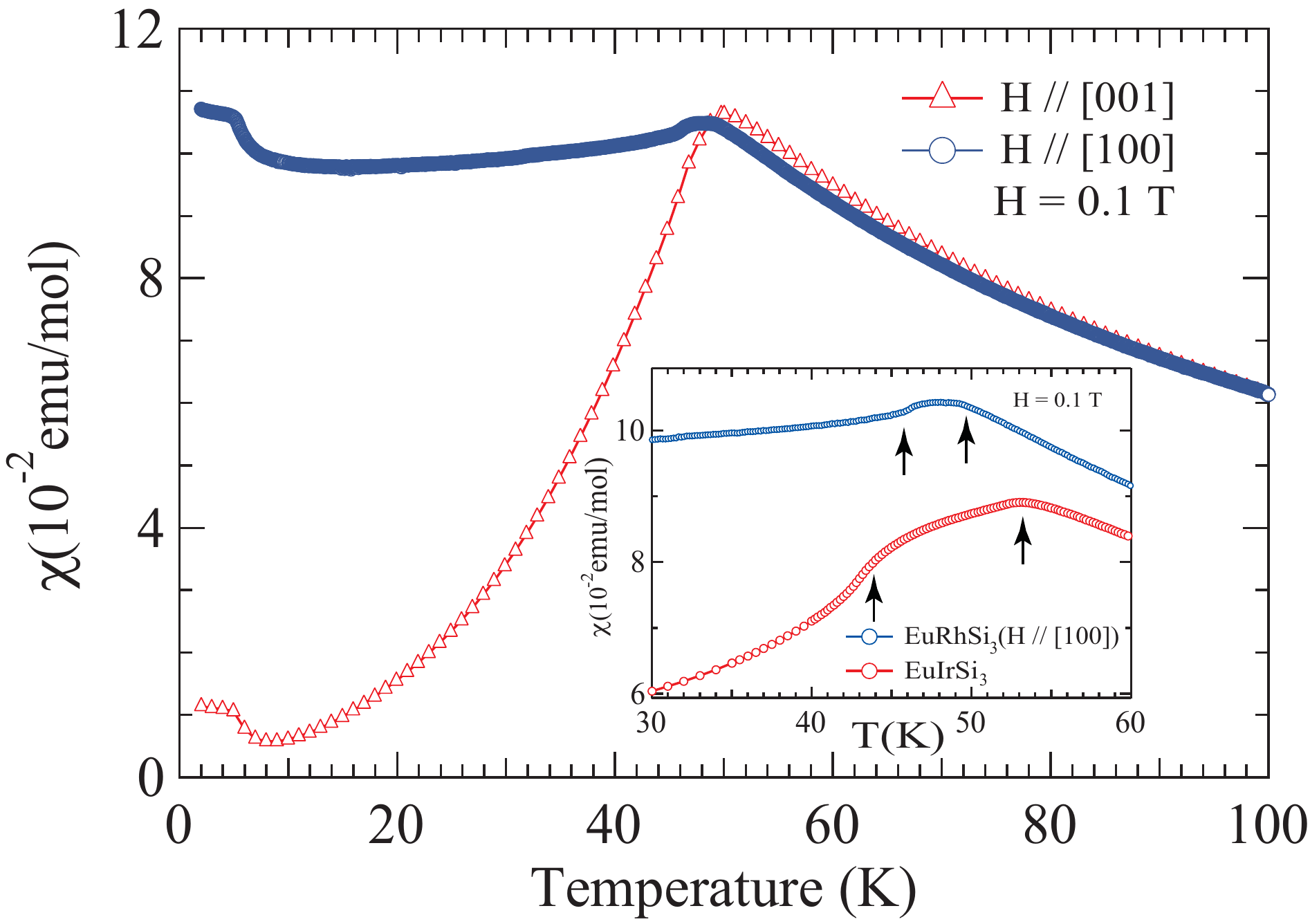}
\caption{\label{Chi_T} Magnetic susceptibility $\chi(T)$ of single crystalline \ers\ with a field of 0.1\,T along [001] and [100]. The slight increase of $\chi$ below 8\,K is due to some parasitic phase, though no unidentified line appears in the x-ray diffraction spectrum. Inset: close-up of $\chi(T)$ for \ers\ with $H~\parallel$~[100] and for polycrystalline \eis.}
\end{figure}
%
The susceptibility of \ers\ measured in a field of 0.1\,T applied along [001] and [100] is shown below 60\,K in Fig.\ref{Chi_T}. It is strongly anisotropic and a clear peak near 49\,K shows the onset of the AF phase. The close-up in the inset of Fig.\ref{Chi_T} shows that this peak is split (arrows). In agreement with the heat capacity and $^{151}$Eu M\"{o}ssbauer data ({\it vide infra}), these two peaks correspond to closely spaced magnetic transitions near 48 and 46\,K. The susceptibility at high temperature (not shown) is nearly isotropic and a fit of the 1/$\chi$ data to a Curie-Weiss law furnishes effective moments $\mu_{\rm eff}$=7.39 and 7.52\,\mub\ and paramagnetic Curie temperature $\theta_{\rm p}=-$11 and $-$14\,K for H along [001] and [100] respectively. These effective moments are lower than the free ion value of 7.94\,\mub\ expected for Eu$^{2+}$ ($g=2$, $S=7/2$), which is due either to the presence of residual In-flux or to a slight Eu off-stoichiometry, corresponding to about 10 at.\% Eu deficit. As to polycrystalline \eis, its susceptibility is shown between 30 and 60\,K in the inset of Fig.\ref{Chi_T}: two anomalies (arrows) are also present, near 52 and 43\,K, witnessing the same phenomenon as in \ers. The $\theta_{\rm p}$ value is $-$17\,K, while $\mu_{\rm eff}$=7.7\,\mub\ is closer to the Eu$^{2+}$free ion value. 
%
\begin{figure}[h]
\centerline{\includegraphics[width=0.37\textwidth]{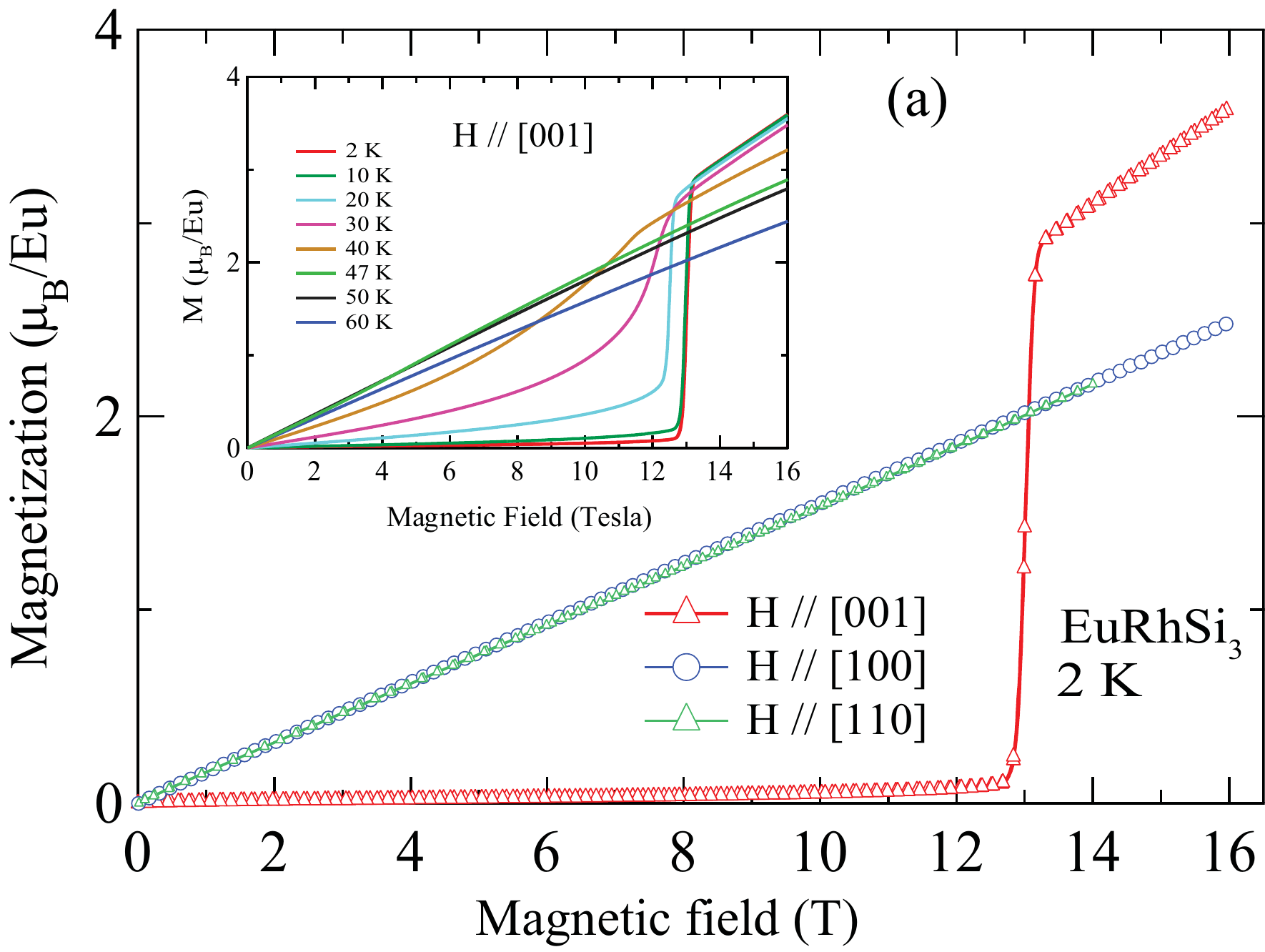}}
\vspace{0.1cm}
\centerline{\includegraphics[width=0.36\textwidth]{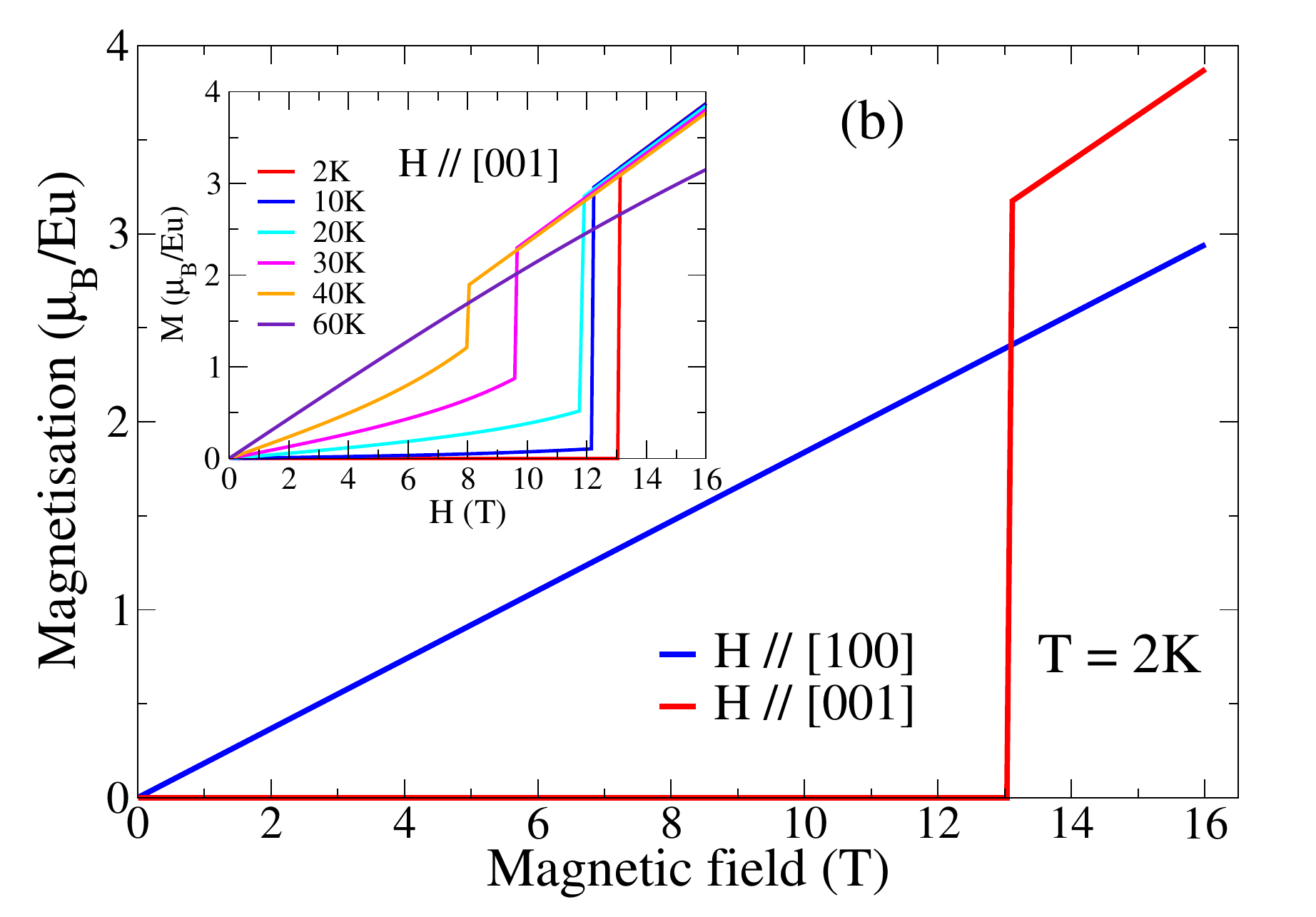}}
\caption{\label{MH} (a) Isothermal magnetisation $M(H)$ of \ers\ at 2\,K along major crystallographic directions. Inset: $M(H)$ curves at higher temperatures. (b) Simulation of the $M(H)$ curves in \ers\ according to the model described in the text. The calculated values are 10\% higher than the data due to the assumed Eu mass deficit in the single crystal sample.}
\end{figure}
%

The N\'eel temperature in intermetallic compounds is the result of the indirect RKKY exchange between $4f$ spins {\bf S} mediated by the $4f$-conduction electron coupling with constant $J_{kf}$: 
\begin{equation}
\label{kf}
{\cal H}_{kf} = -J_{kf}\  {\bf s}.{\bf S}, 
\end{equation}
where {\bf s} is the conduction electron spin density, and it has the form: $T_{\rm N} \propto J_{kf}^2\  n(E_F)\  S(S+1)$ \cite{freeman}. Its value in these materials, near 50\,K, is much larger than in other members of the series ($\simeq 15$\,K), pointing to an enhanced value of $n(E_F)$.
  
The isothermal magnetisation versus field scan at 2\,K in \ers, shown in Fig.\ref{MH}(a), is a textbook example of an antiferromagnet with the tetragonal [001] c-axis as the easy axis of magnetisation and the (001) $ab$-plane as the hard plane, in line with the susceptibility data of Fig.\ref{Chi_T}. For $H~\parallel$~[001], an unusually large spin-flop field of 13\,T is observed. A linear extension of the magnetisation curves for both field directions up to a saturation moment value $m_0$= 7\,\mub\ yields large spin-flip fields H$_{sf}^c \simeq$28\,T and H$_{sf}^a \simeq$41\,T. The latter is larger since the field aligns the moments in the hard magnetic plane. In the standard molecular field theory \cite{Herpin}, one has: H$_{sf}^c$=2(H$_e -$H$_a$) and H$_{sf}^a$=2(H$_e +$H$_a$), where H$_e$ and H$_a$ are respectively the exchange and the ``anisotropy'' field, the latter being defined as H$_a=K/m_0$ where $K$ is the anisotropy energy density. Then we obtain H$_e$=17\,T and H$_a$=3.35\,T, and the critical spin-flop field H$_{cr}$ = 2$\sqrt{{\rm H}_a({\rm H}_e-{\rm H}_a)}$ = 13.5\,T, in very good agreement with experiment. The spin-flop field in \ers\ is much larger than in EuNiGe$_3$ \cite{Arvind_EuNiGe3} (2-3\,T), a logical consequence of a large H$_e$ linked to the high $T_{\rm N}$ value. 

In the inset of Fig.\ref{MH}(a) are plotted the magnetisation curves along the easy axis at higher temperature. On heating, the spin-flop field decreases and the jump at the spin-flop broadens.

\subsection{Modeling}

However, such a simple model as described above cannot account for the quite different values of $T_{\rm N}$ and $\vert \theta_p \vert$, and the oscillatory nature of the RKKY exchange compels one to introduce at least two different exchange integrals. Therefore, we use a numerical self-consistent calculation which has been described thoroughly in Ref.\cite{Arvind_EuNiGe3}: i) the infinite range dipolar interaction is added, ii) two exchange integrals $J_1$ (intra-plane first neighbor) and $J_2$ (interplane first neighbor) for the centered tetragonal structure are considered, as well as exchange anisotropy, and iii) the single ion crystalline anisotropy is described by a term $D S_z^2$, where $Oz$=c. We also assume a magnetic structure made of ferromagnetic (ab) planes ($J_1 > 0$) coupled antiferromagnetically along c ($J_2 < 0$), i.e. a propagation vector {\bf k}=[001]. To obtain a first estimation of the exchange constants, the molecular field equations linking $T_{\rm N}$ and $\theta_p$ with $J_a$ and $J_c$ are used \cite{Johnston_EuNiGe3,Arvind_EuNiGe3}, yielding: $J_a$=1.1\,K and $J_c=-$0.88\,K. We have taken $T_{\rm N}$=60\,K (see the section about M\"ossbauer spectroscopy) and a mean value $\theta_p=-$13\,K. Figure \ref{MH}(b) shows the curves which reproduce best the experimental data, with a small exchange anisotropy: $J_a^{\parallel}$=0.8\,K, $J_a^\perp$=1.1\,K, $J_c^{\parallel}=-$0.7\,K, $J_c^\perp=-$0.9\,K and a crystalline anisotropy parameter $D=-$0.85\,K. The model yields $T_{\rm N}$=57\,K and $\theta_p$=$-$6 and $-$17\,K for $H~\parallel$~[001] and $H~\parallel$~[100] respectively (experimental values $-$11 and $-$14\,K resp.). At 30\,K and above, the model cannot exactly reproduce the smoothing of the spin-flop transition (see inset of Fig.\ref{MH}(b)).

\subsection{Electrical resistivity}

\begin{figure}[h]
\centerline{\includegraphics[width=0.35\textwidth]{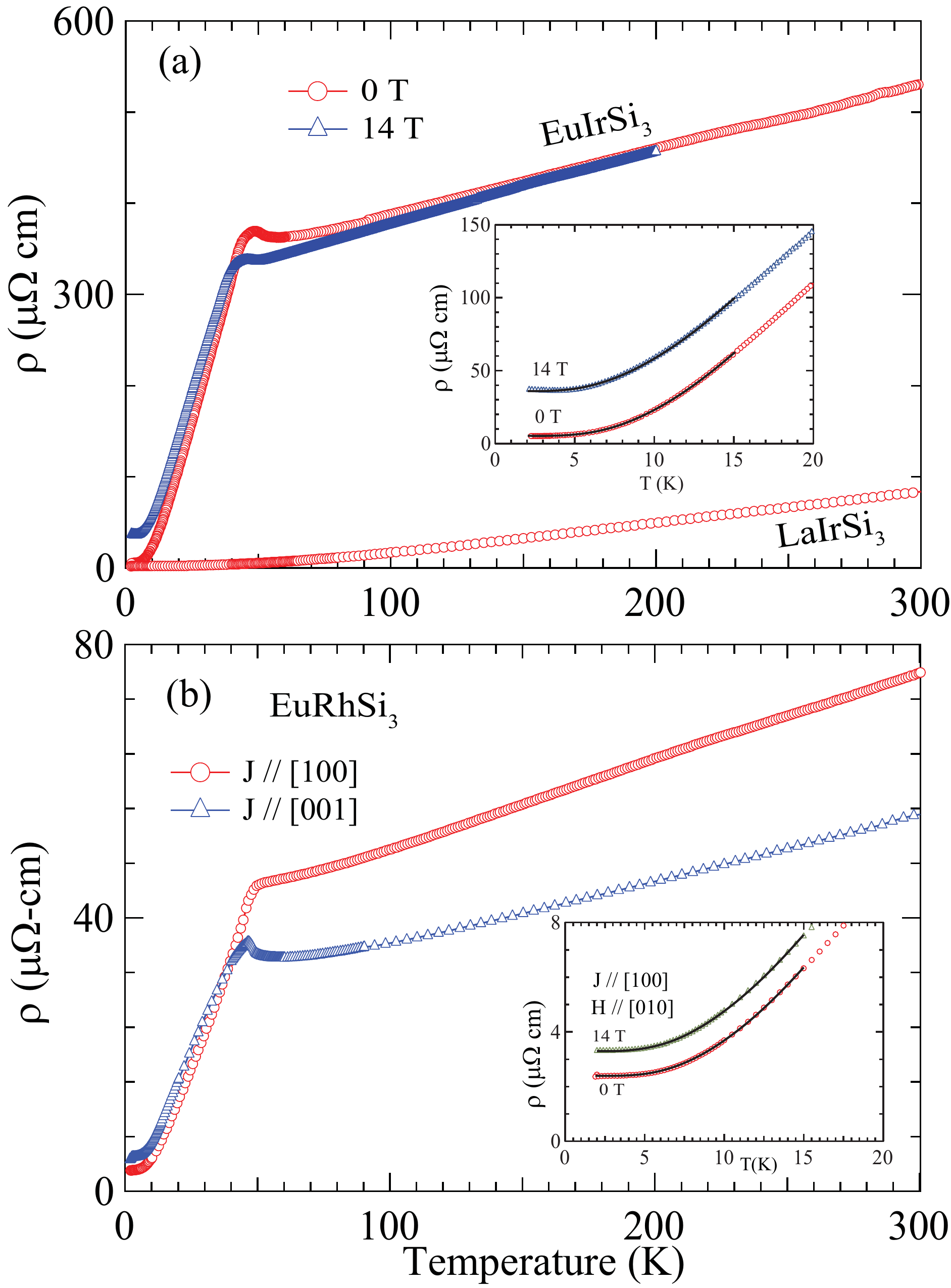}}
\caption{\label{RT} (a) Electrical resistivity as a function of temperature $\rho(T)$ of EuIrSi$_3$ in zero field and in a 14\,T field. (b) $\rho(T)$ data of EuRhSi$_3$ when current density $J~\parallel$~[100] and [001]. The insets show the low temperature data together with their fit to Eq.(\ref{rho_gap}).}
\end{figure}

The electrical resistivity of \eis\ and \ers\ (and of non-magnetic LaIrSi$_3$) is shown in Figs.\ref{RT}(a) and (b). Above T$_N$ it varies almost linearly up to 300\,K in both compounds due to phonon scattering. It shows a small upturn at the magnetic transitions, likely caused by antiferromagnetic fluctuations just above $T_{\rm N1}$ and decreases rapidly on cooling due to a strong depletion in the spin-disorder scattering in the magnetically ordered state. One striking feature is that the resistivity of polycrystalline \eis\ is one order of magnitude larger than that of single crystal \ers. We reckon that this is probably due to strong grain boundary scattering which enhances the resistivity of the polycrystalline sample. The residual resistivity $\rho_0=\rho(2\,K)$ is lower than 10\,$\mu$Ohm.cm in both materials (see Table \ref{Table1}), indicating they are chemically well ordered. 
%

The total spin disorder resistivity $\rho_m$, estimated as $\rho(T_{\rm N}) - \rho_0$, amounts to about 35\,$\mu$Ohm.cm in \ers\ while it is of the order of a few $\mu$Ohm.cm in the other compounds of the series \cite{Neeraj2010EuPtSi3,Neeraj2012EuPtGe3,Arvind_EuNiGe3}. Since $\rho_m \propto J_{kf}^2\ n(E_F)\ S(S+1)$ \cite{freeman}, its enhanced value in \ers\ is in line with a stronger $n(E_F)$.

In the AF phase below 20\,K, magnon scattering is the dominant mechanism for resistivity. With an antiferromagnetic spin wave  dispersion relation:  
\begin{equation}
\label{disp}
E(k)=\sqrt{\Delta^2+\sigma k^2}
\end{equation} 
where $\Delta$ is the anisotropy gap and $\sigma$ the spin-wave stiffness, the electrical resistivity in zero field for $T < \Delta$ is given by \cite{fontes}: 
\begin{eqnarray}
\label{rho_gap}
\rho (T)=\rho_0 &+& A \Delta^2 {\left(\frac{T}{\Delta}\right)}^{1/2}e^{- \Delta/T} \\ \nonumber
& \times &\left[1+2/3 \left(\frac{T}{\Delta}\right)+2/15\left(\frac{T}{\Delta}\right)^2\right],
\end{eqnarray}
\begin{table}[!]
\caption{\label{Table1} Parameters obtained after fitting Eq.(\ref{rho_gap}) to the zero field resistivity data of \ers\, with $J \parallel [001]$, and of \eis.}
\begin{tabular}{|c|c|c|c|}   \hline \hline
        & $\rho_0(\mu\Omega~cm$) & $A(\mu\Omega~cm/K^2$) & $\Delta (K)$ \\ \hline
\ers \  & 5.1  & 0.04 & 23.3 \\ \hline
\eis    & 5.27  &  0.43  &  25.2 \\ \hline
\end{tabular} 
\end{table}
%
where $A$ a material dependent constant. Equation (\ref{rho_gap}) provides a good fit to the zero field data in both compounds between 1.8 and 15\,K (and also for a 14\,T field), as shown by the solid lines in the insets of Fig.\ref{RT}. In \eis, the zero field spin wave gap is 25.2\,K and it is 23.5\,K in \ers. The AF spin wave gap is linked to the exchange and anisotropy fields H$_e$ and H$_a$ by the relation \cite{tqs}:
\begin{equation}
\label{gap_h}
\Delta = 2g \mu_B \sqrt{{\rm H}_a\ ({\rm H}_a+{\rm H}_e)}.
\end{equation}
Using the values determined above in \ers: H$_e$=17\,T, H$_a$=3.35\,T and $g$=2, one obtains $\Delta$ = 22.1\,K, in excellent agreement with experiment in this material and close to the value in \eis.

At a field of 14\,T, $\rho_0$ increases to 35.9\,$\mu\Omega$cm in \eis\ and to 9\,$\mu\Omega$cm in \ers\ for $H~\parallel$~[100] and $J~\parallel$~[001]. 

\begin{figure}[h]
\centerline{\includegraphics[width=0.40\textwidth]{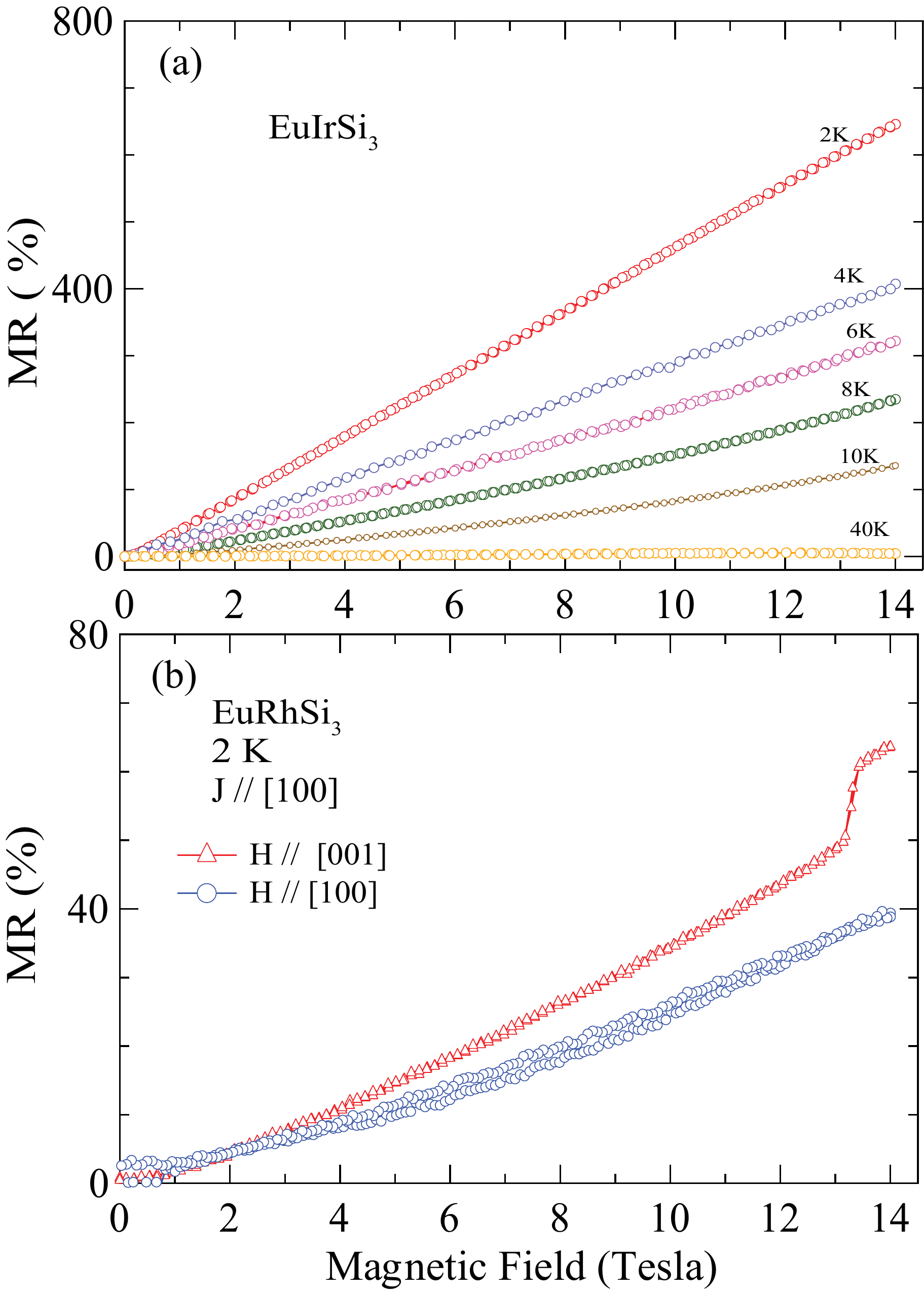}}
\caption{\label{MR} Magnetoresistance ($MR$) of EuIrSi$_3$ at different temperatures. (b) $MR$ of EuRhSi$_3$ for $H~\parallel$~[100] and $H~\parallel$~[001] with $J~\parallel$~[100] in both the cases.}
\end{figure}
%
The significant increase of $\rho_0$ with field corresponds to a positive magnetoresistance ratio $MR$, defined as $MR=[\rho(H)-\rho(H=0)]/\rho(H=0)$, as expected for antiferromagnets \cite{yamada,mcewen}. The plots of $MR$ at selected temperatures in \eis\ are shown in Fig.\ref{MR}(a). It reaches giant values: at 14\,T, it exceeds 100\% at 10\,K and it reaches 600\% at 2\,K. At higher temperatures (not shown), $MR$ decreases and becomes negative near 47\,K, shows an absolute minimum near 50\,K and has a small residual value of $-$3\% at 200\,K, far above $T_{\rm N1}$. 
In \ers, the $MR$ is lower, as shown in Fig.\ref{MR}(b) at 2\,K for $J~\parallel$~[100] and $H~\parallel$~[001] and [100]. A small positive jump in the $MR$ occurs for $H~\parallel$~[001] at the spin-flop field (13\,T), although the $MR$ is expected to drop to a small (positive) value above the spin-flop field, according to the molecular field calculation in Ref.\cite{yamada}.
\subsection{Heat capacity}
\begin{figure}[h]
\centerline{\includegraphics[width=0.37\textwidth]{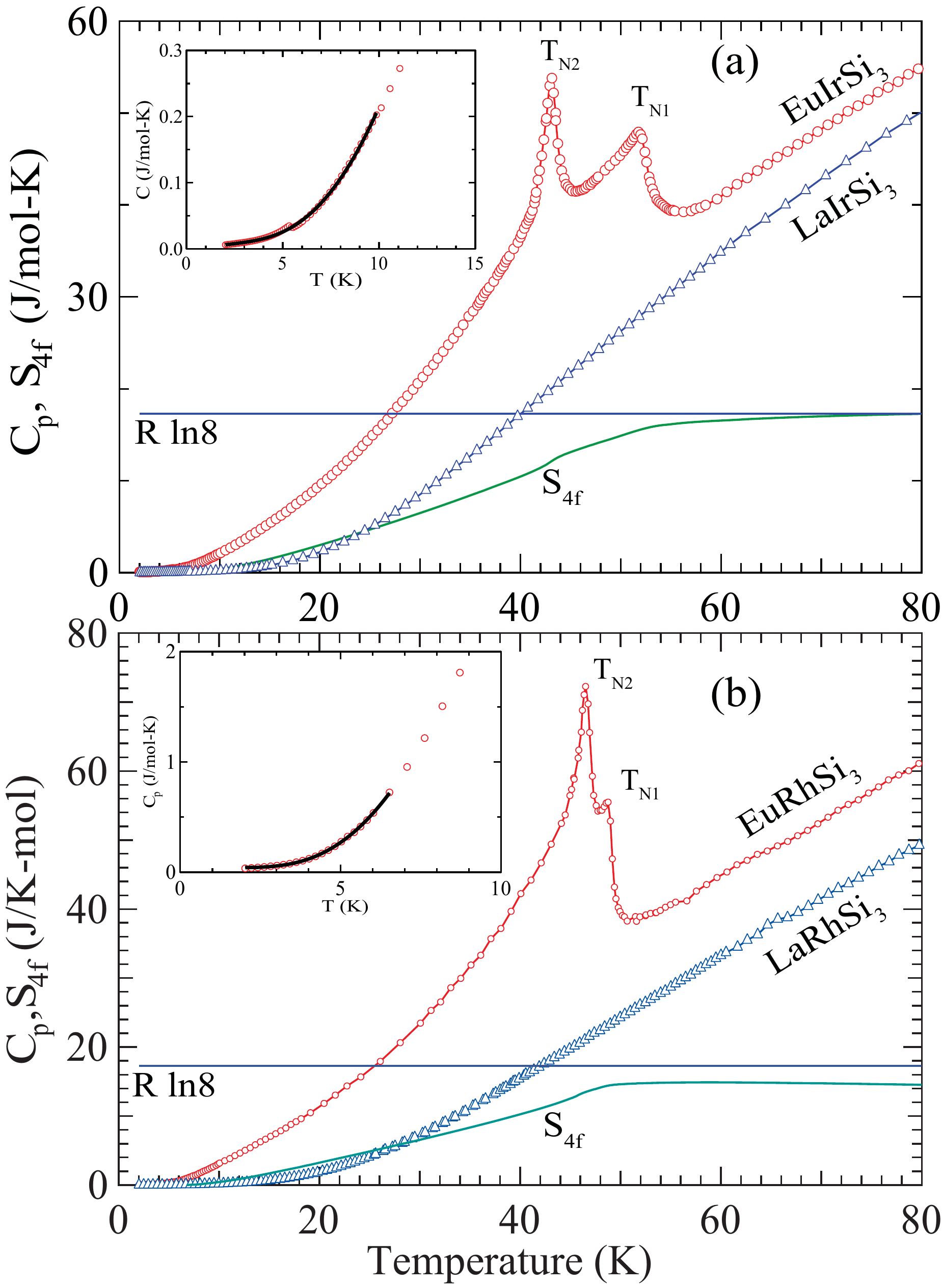}}
\caption{\label{HC} (a) Variation of heat capacity with temperature $C(T)$ of \eis\ and of its non magnetic analogue LaIrSi$_3$. The solid green line represents the calculated entropy S$_{4f}(T)$. Inset shows the fitted Eq.(\ref{cp}) to the the low temperature data (b) $C(T)$ of \ers\ and LaRhSi$_3$ with a similar legend as for the upper panel.}
\end{figure}

The main panels of Fig.\ref{HC} show the heat capacity of \eis\ and \ers\ together with that of the non-magnetic La-reference. The two plots show two major peaks at 51.8 and 43.1\,K in \eis\ and at 48.3 and 45.8\,K in \ers\ in close correspondence with the anomalies seen in the susceptibility data, thus confirming  the occurrence of two magnetic transitions in these two compounds. The jump in the heat capacity at the higher transition temperature $T_{\rm N1}$ is approximately 9.5 and 12\,J/mol.K in the Ir and Rh-compound, respectively, which is lower than the value $\delta C_{7/2}$=20.14\,J/mol.K for an equal moment antiferromagnetic transition for $S$=7/2 ions in the mean field model, and closer to that predicted for an amplitude modulated structure (2/3 $\delta C_{7/2}$) \cite{Neeraj2010EuPtSi3,gignoux}. This suggests that at $T_{\rm N1}$ occurs a transition to a modulated moment structure. The transition from this intermediate structure to an equal moment structure takes place at $T_{\rm N2}$. This behaviour is confirmed by the $^{151}$Eu M\"osbauer spectra recorded at few selected temperatures. 

Deep in the AF phase, the specific heat should be the sum of 3 terms \cite{conti}: 
%
\begin{eqnarray}
\label{cp}
C(T)~=~\gamma T &+& \beta T^3 + B \Delta^4 {\left(\frac{T} {\Delta}\right)}^{1/2} e^{- \Delta/T} \\ \nonumber &\times& \left[1+39/20 \left(\frac{T}{\Delta}\right)+51/32\left(\frac{T}{\Delta}\right)^2\right],
\end{eqnarray}
%
where the first linear term is the conduction electron heat capacity, the second is the phonon contribution and the third the magnon heat capacity corresponding to a dispersion law given by Eq.(\ref{disp}) with a gap $\Delta$. 
Equation (\ref{cp}) provides a good fit of the data between 1.8 and 10\,K, shown by the solid line in the insets of Figs.\ref{HC}(a) and (b). For \eis, the best fit estimates of $\gamma$ and $\Delta$ are 30.5\,mJ/mol.K$^2$ and 22.9\,K respectively. The gap value is very close to that inferred from the resistivity data (25.2\,K), which lends credibility to our analysis. In the case of \ers, one obtains: $\gamma$= 40.3\,mJ/mol.K$^2$ and $\Delta$=17.6\,K, the latter value somewhat lower than the resistivity derived value (23.3\,K). It may be noted that the Sommerfeld coefficient $\gamma$ is an order of magnitude larger than in $sp$-metals, and even larger than in many $d$-metal alloys. The Sommerfeld coefficient is expressed as \cite{tqs}: $\gamma = \frac{2}{3} \pi^2 k_B^2\ n(E_F)$, and its large value must be related to an enhanced $n(E_F)$. 
%
	
The entropy $S_{4f}$ associated with the magnetic ordering was estimated by integrating $C_{4f}/T$, where $C_{4f}$ was obtained by subtraction from $C(T)$ of both the normalized heat capacity of LaTSi$_3$, taking into account the slight difference in the atomic masses of Eu and La, and the large conduction electron contribution $\gamma T$. It is seen that $S_{4f}$ attains the expected value of $R \ln8$ (for $S$~=~7/2 Eu$^{2+}$ ions) close to $T_{\rm N1}$ in both compounds.

\subsection{$^{151}$Eu M\"{o}ssbauer spectra}  
 
\begin{figure}[h]
\centerline{\includegraphics[width=0.47\textwidth]{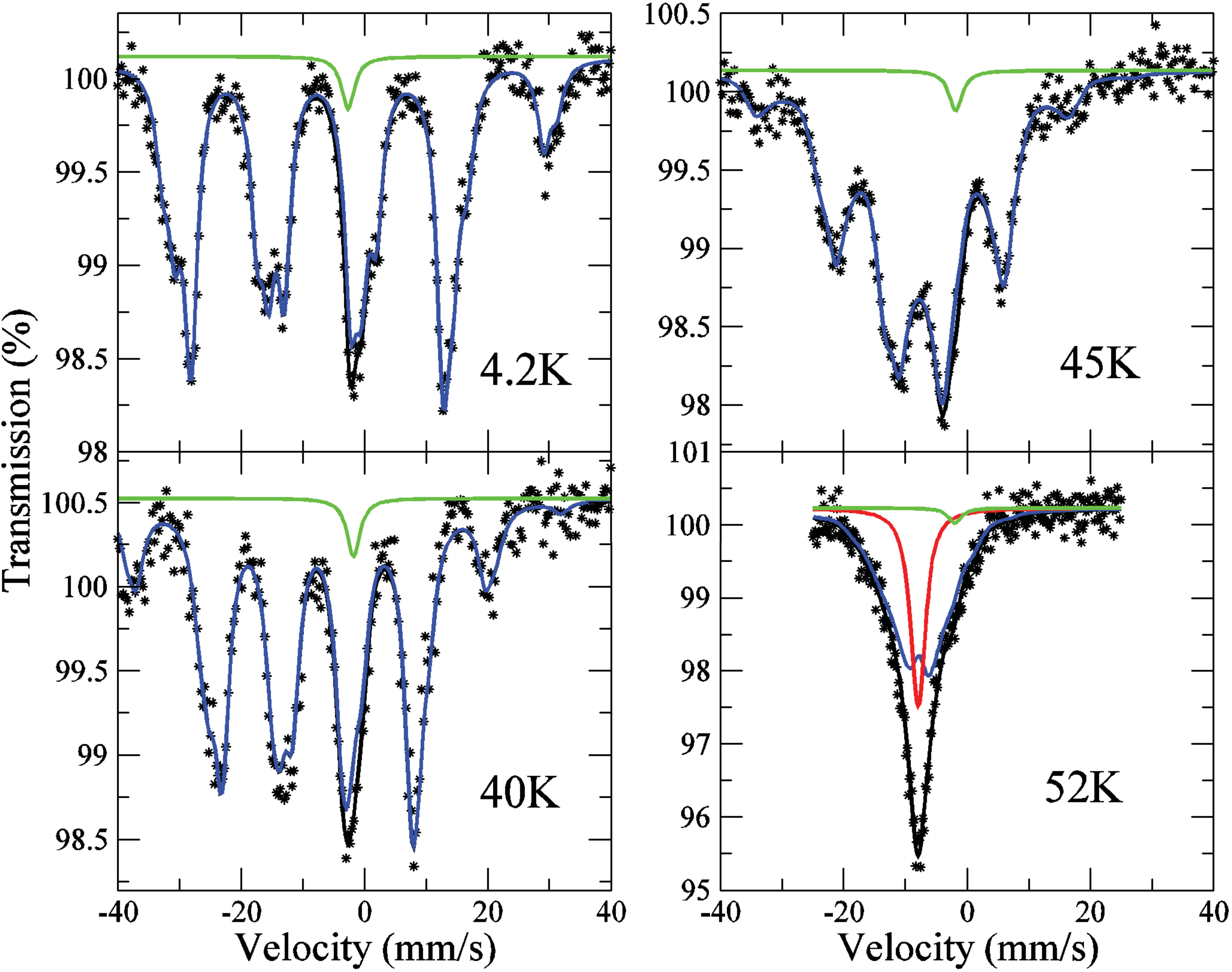}}
\caption{\label{Moss1} $^{151}$Eu M\"ossbauer spectra in \eis\ at selected temperatures: in the commensurate phase (4.2 and 40\,K), and in the incommensurate phase (45 and 52\,K). The spectrum at 52\,K is typical of the ``coexistence'' region, with a contribution from the paramagnetic phase of some part of the sample (red subspectrum) occurring in case of a first order transition. The green subspectrum represents an impurity phase containing Eu$^{3+}$ (relative intensity 2\%).}
\end{figure}

The M\"ossbauer spectra on the isotope $^{151}$Eu in \eis\ are shown in Fig.\ref{Moss1} in the two temperature ranges defined as phase I ($T <$43.1\,K) and phase II (43.1\,K $<$ T $<$ 51.8\,K). Spectra in \ers\ are similar. The spectrum at 4.2\,K is a standard hyperfine field pattern characteristic of Eu$^{2+}$ ($S$=7/2, $L$=0), with an isomer shift relative to EuF$_3$ of $-$7.92(5)\,mm/s which matches well with the value reported in Ref.\cite{Chevalier}. This spectrum presents the peculiarity of a very high hyperfine field H$_{hf}$ (4.2\,K) = 52.6(3)\,T (51.6\,T in \ers). This high value can be explained by the large $n(E_F)$ prevailing in these materials. Indeed, in magnetically ordered intermetallic materials with the $L$=0 ion Eu$^{2+}$, the hyperfine field is solely due to the spin polarisation of the $s$-type electrons at the nucleus site. It can be expressed as \cite{bleaney}: $
{\rm H}_{hf}(T) = A \ m(T) + {\rm H}_{ce}$, where the first term is the core polarisation field proportional to the Eu$^{2+}$ moment $m(T)=-g \mu_B \langle S \rangle_T$ and worth $\simeq 34$\,T at saturation \cite{nowik}. The second term H$_{ce}$ is due to the conduction electron spin polarisation $\langle s \rangle_T$ induced by the $4f$ shell through $J_{kf}$ exchange and is given approximately by \cite{bleaney}: $
{\rm H}_{ce} \simeq A_{ce} \ \mu_B \ \langle s \rangle_T $, where $A_{ce}$ is a hyperfine constant. Acording to Eqn.(\ref{kf}), the effective field on {\bf s} is: $\frac{J_{kf} {\bf S}}{g\mu_B}$, and introducing the Pauli susceptibility: $\chi_P = 2 \ \mu_B^2\  n(e_F)$, one obtains:
\begin{equation}
\label{hce}
{\rm H}_{ce} \simeq A_{ce} \ J_{kf} n(E_F)\ m(T).
\end{equation}
In most Eu$^{2+}$ intermetallic materials and in the other members of the series, the saturated hyperfine field amounts to about 30\,T, i.e. H$_{ce}$ is negative and worth a few T. In \eis\ and \ers, due to the enhanced $n(E_F)$ value, H$_{ce}$ is much larger ($\simeq$18\,T) and happens to be positive. The same situation holds in EuFe$_4$P$_{12}$, which presents the largest hyperfine field ever measured with $^{151}$Eu (67\,T), implying H$_{ce} \simeq$33\,T \cite{gerard}.

\begin{figure}[h]
\centerline{\includegraphics[width=0.37\textwidth]{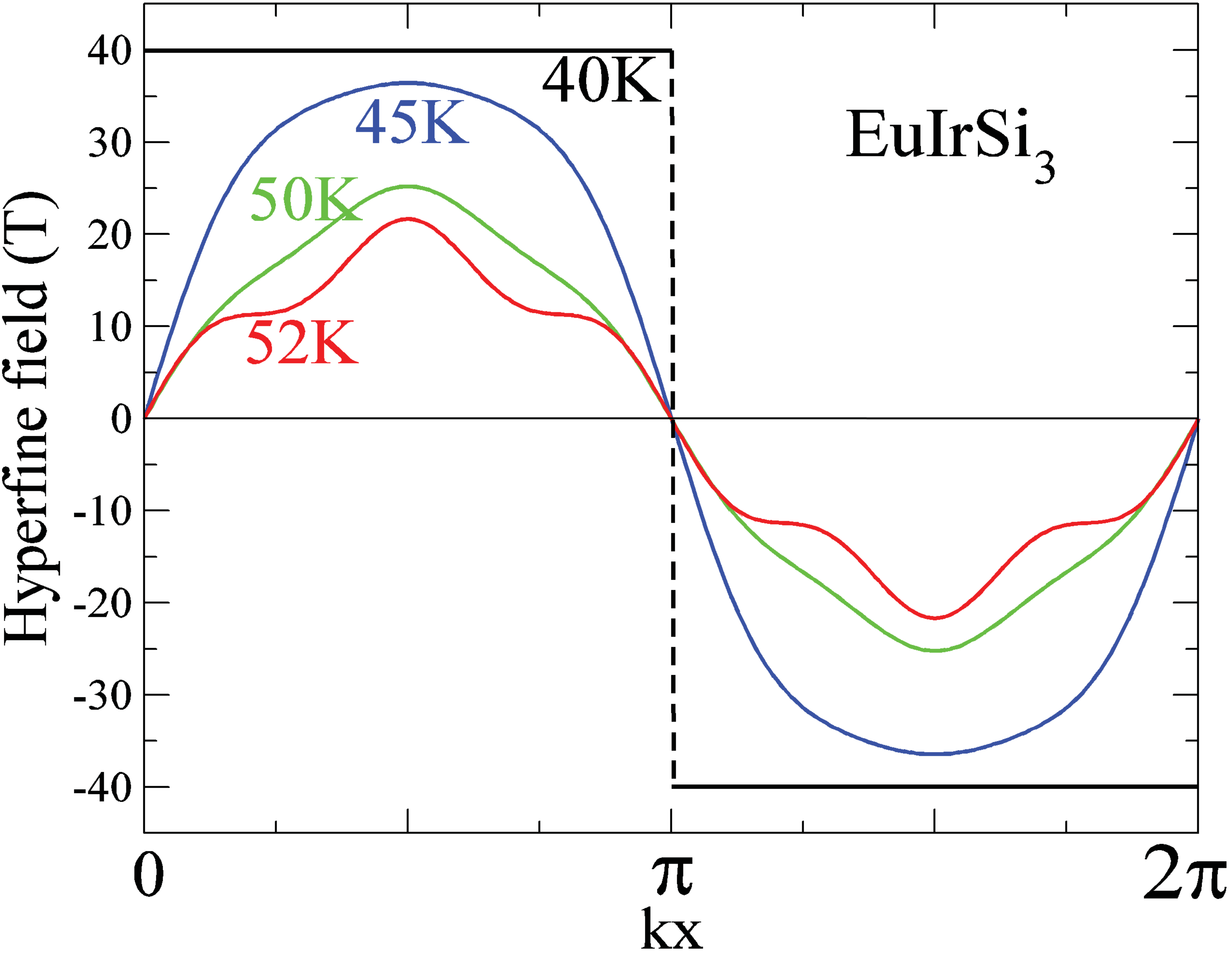}}
\caption{\label{Moss2} Hyperfine field modulation over a period in \eis\ in the incommensurate modulated phase II at selected temperatures.}
\end{figure}

Whereas the spectra in phase I (below 43\,K) show the presence of a unique hyperfine field, and hence of a unique Eu$^{2+}$ magnetic moment, there occurs a sudden change of the spectral shape above 43\,K, i.e. when entering phase II, where the spectra can be fitted to an incommensurate modulation of hyperfine fields \cite{bonville}. The latter is well described by the first 3 odd harmonics of a Fourier series:
\begin{equation}
\label{eq.4}
H_{hf}(kx)=h_1 \sin(kx)+h_3 \sin(3kx)+h_5 \sin(5kx)
\end{equation} 
where $k$ is the propagation vector of the modulation and $x$ the distance along $k$. The three coefficients $h_1$, $h_3$ and $h_5$ were fitted to the spectral shape at each temperature to obtain the modulation profiles shown in Fig.\ref{Moss2}. The modulation becomes more ``squared'' as temperature decreases and approaches the transition to the incommensurate phase.

\begin{figure}[h]
\centerline{\includegraphics[width=0.37\textwidth]{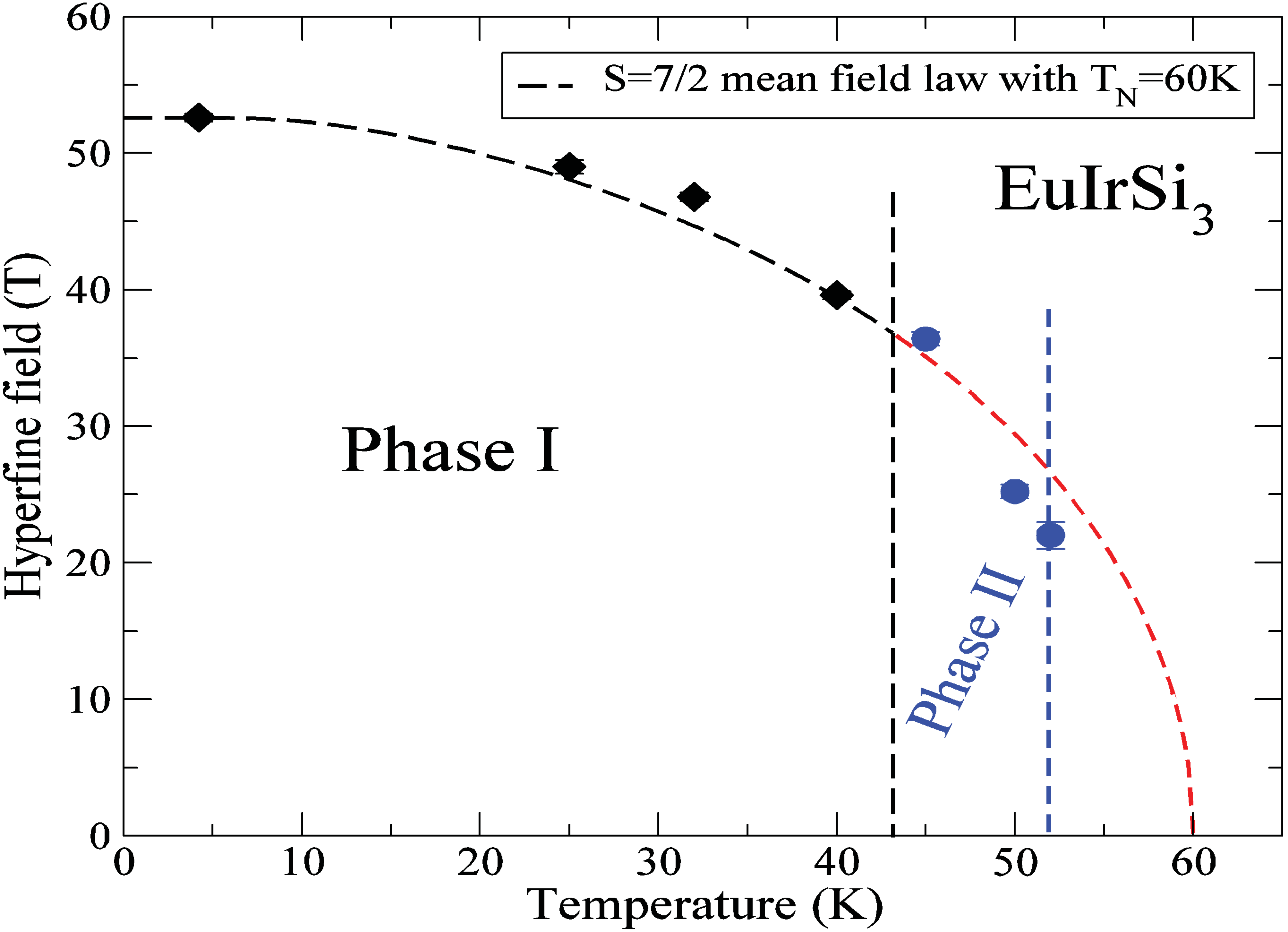}}
\caption{\label{Moss3} In \eis, thermal variation of the hyperfine field  in Phase I (black dots) and of the maximum of the hyperfine field modulation in Phase II (blue disks). The dashed line is the $S$=7/2 mean field law with $T_{\rm N}$=60\,K, extrapolated in phase II above 43\,K (in red).}
\end{figure}
The thermal variation of the hyperfine field is plotted in Fig.\ref{Moss3}. In phase I, the hyperfine field values approximately follow a mean field law for $S$=7/2, in line with its proportionality to the Eu$^{2+}$ moment derived above. The transition temperature of this mean field law (60\,K) does not correspond to the actual N\'eel temperature because of the presence of the commensurate - incommensurate transition at 43.1\,K and of the first-order character of the transition to the paramagnetic phase at 51.8\,K (see spectrum at 52\,K in Fig.\ref{Moss1}).

\section{Conclusion}.

The whole set of our thermodynamic and spectroscopic measurements in the two divalent Eu intermetallics \eis\ and \ers\ can be coherently and qualitatively interpreted by assuming a high density of electronic band states at the Fermi energy, which sets them apart from the other members of the EuTX$_3$ family. We attribute the observed enhanced values to a large $n(E_F)$ rather than to an anomalously large $4f$-conduction electron coupling $J_{kf}$ since the Sommerfeld coefficient does not involve $J_{kf}$ and there is {\it a priori} no reason for the stable Eu$^{2+}$ $4f$ shell to be prone to strong hybridisation with the conduction band. The two compounds present a cascade of magnetic transitions near 50\,K, from a paramagnetic to an incommensurate modulated, then to a commensurate antiferromagnetic phase. In the \ers\ single crystal sample, we could evidence an important anisotropy of the magnetisation, confirming the link between these two phenomena.   


\begin{thebibliography}{0}
\bibitem{freeman}
\Name{see for instance Freeman A. J.}
\Book{Magnetic properties of rare earth metals, Chap.6}
\Editor{Elliott R. J., Plenum Press, London and New York, 1972.}

\bibitem{Neeraj2010EuPtSi3}
\Name{Kumar N., Dhar S. K., Thamizhavel A., Bonville P. \and Manfrinetti P.}
\REVIEW{Phys. Rev. B}{81}{2010}{144414}.
  
\bibitem{Neeraj2012EuPtGe3}
\Name{Kumar N., Das P. K., Kulkarni R., Thamizhavel A., Dhar S. K. \and Bonville P.}
\REVIEW{J. Phys.: Condens. Matter}{24}{2012}{036005}.
  
\bibitem{EuPdGe3}
\Name{Bednarchuk O., Gagor A. \and Kaczorowski D.}
\REVIEW{J. Alloys Comp.}{622}{2015}{432}.

\bibitem{Johnston_EuNiGe3}
\Name{Goetsch R. J., Anand V. K. \and Johnston D. C.}
\REVIEW{Phys. Rev. B}{87}{2013}{064406}.

\bibitem{Arvind_EuNiGe3}
\Name{Maurya A., Bonville P., Thamizhavel A. \and Dhar S. K.}
\REVIEW{J. Phys.: Condens. Matter}{87}{2013}{064406}.

\bibitem{Chevalier}
\Name{Chevalier B., Coey J. M. D., Lioret B. \and Etourneau J.}
\REVIEW{J. Phys. C: Solid State Phys.}{19}{1986}{4521}.

\bibitem{Herpin}
\Name{Herpin A.}
\Book{Th\'eorie du Magn\'etisme}
\Editor{Presses Universitaires de France, Paris, France}{(1968)}

\bibitem{fontes}
\Name{Fontes M. B., Trochez J. C., Giordanengo B., Bud'ko S. L., Sanchez D. R., Baggio-Saitovitch E. M. \and Continentino M. A.}
\REVIEW{Phys. Rev. B}{60}{1999}{6781}

\bibitem{tqs}
\Name{Kittel C.}
\Book{Quantum theory of solids}
\Editor{John Wiley and Sons, New-York}{(1964)}

\bibitem{yamada}
\Name{Yamada H. \and Takada S.}
\REVIEW{Progr. Theor. Phys.}{49}{1973}{1401}

\bibitem{mcewen}
\Name{McEwen K. A.}
\Book{Handbook on the Physics and Chemistry of rare Earths, p.411}
\Editor{Gschneider K. A. \and Eyring L., North-Holland, Amsterdam,1978}

\bibitem{gignoux}
\Name{Blanco J. A., Gignoux D. \and Schmitt D.}
\REVIEW{Phys. Rev. B}{43}{1991}{13145}

\bibitem{conti}
\Name{Continentino M. A., de Medeiros S. N.,Orlando M. T. D., Fontes M. B. \and Baggio-Saitovitch E. M.}
\REVIEW{Phys. Rev. B}{64}{2001}{012404}

\bibitem{bleaney}
\Name{Bleaney B.}
\Book{Magnetic properties of rare earth metals, Chap.8}
\Editor{Elliott R. J., Plenum Press, London and New York, 1972}

\bibitem{nowik}
\Name{Nowik I., Dunlap B. D. \and Wernick J. H.}
\REVIEW{Phys. Rev. B}{8}{1973}{238}

\bibitem{gerard}
\Name{G\'erard A., Grandjean F., Hodges J. A., Braun D. J. \and Jeitschko W.}
\REVIEW{J. Phys. C: Solid State Phys.}{16}{1983}{2797}

\bibitem{bonville}
\Name{Bonville P., Hodges J. A., Shirakawa M., Kasaya M. \and Schmitt D.}
\REVIEW{Eur. Phys. J. B}{21}{2001}{349}

\end{thebibliography}
\end{document}